\documentclass[journal=jacsat,manuscript=article]{achemso}
\usepackage[version=3]{mhchem}
\usepackage{color}

\title[]{Superconductivity in W$_{3}$Re$_{2}$C with chiral structure}
\author{Lei Yang$^{1,2,\dag}$, Jing Jiang$^{1,2,\dag}$, Hui-Hui He$^{1,2,\ddag}$, Ying Ma$^{3}$, Kai Liu$^{1,2,*}$, Xiao Zhang$^{3,*}$, and Hechang Lei$^{1,2,}$}
\email{kliu@ruc.edu.cn, zhangxiaobupt@bupt.edu.cn, hlei@ruc.edu.cn}
\affiliation[RUC]
{$^{1}$School of Physics and Beijing Key Laboratory of Opto-electronic Functional Materials $\&$ Micro-nano Devices, Renmin University of China, Beijing 100872, China\\
$^{2}$Key Laboratory of Quantum State Construction and Manipulation (Ministry of Education), Renmin University of China, Beijing 100872, China\\
$^{3}$State Key Laboratory of Information Photonics and Optical Communications $\&$ School of Physical Science and Technology, Beijing University of Posts and Telecommunications, Beijing 100876, China}

\begin{document}
	
\begin{abstract}
We discover superconductivity in cubic W$_{3}$Re$_{2}$C with chiral structure and the superconducting transition temperature $T_{c}$ is about 6.2 K. Detailed characterizations and analysis indicate that W$_{3}$Re$_{2}$C is a bulk type-II BCS superconductor with full isotropic gap. 
Moreover, first-principles calculations indicate that the electron-phonon coupling primarily arises from interactions between W/Re 5$d$ electronic states and their low-frequency phonons. Furthermore, the breaking of inversion symmetry in W$_{3}$Re$_{2}$C facilitates the emergence of Weyl points in the electronic structure. Therefore, W$_{3}$Re$_{2}$C can serve as a promising platform for investigating the influences of chiral structure on both superconductivity and band topology.
\end{abstract}

\newpage

\section{Introduction}
Noncentrosymmetric superconductors (NCSs) have received intensive interest because antisymmetric spin-orbit coupling (ASOC) can break parity conservation, permitting the coexistence and mixing of spin-singlet and spin-triplet components during Cooper pair formation  \cite{Sigrist}.
This mixing can induce some exotic phenomena, such as the emergence of nodes in the superconducting (SC) gap function \cite{Bonalde2}. 
As a unique category of NCSs, due to the absence of improper rotational symmetry elements—including mirror planes and inversion centers—while preserving only proper rotational axes, superconductors with chiral structures exhibit a series of anomalous physical properties, such as nonreciprocal superconducting behaviors and other parity-violating phenomena arising from spin-momentum locking in chiral structures \cite{Sigrist}.
It is predicted that they can host Majorana fermions within their vortex cores and along edges \cite{Volovik,Read,Stone}. These quasiparticles exhibit non-Abelian statistics, which confers topological robustness (intrinsic noise immunity) and thus becomes an ideal building block for quantum computation \cite{Kitaev}.
	
Research on NCSs was greatly stimulated by the discovery of  CePt$_{3}$Si \cite{Samokhin,Bonalde,Mukuda}, a heavy-fermion compound exhibiting coexistence of antiferromagnetism and superconductivity. Its lack of inversion symmetry (space group $P4mm$) induces a strong ASOC, resulting in a mixed singlet-triplet paired state.  
	This manifests in notable properties including nodal superconducting gaps, and an upper critical field that significantly surpasses the Pauli limit. 
	Li$_{2}$Pt$_{3}$B \cite{Bose,Chandra,Lee,Takeya,Yuan,Badica,Miclea,Mukherjee} and  Mo$_{3}$Al$_{2}$C \cite{Koyama,Ozaki,Reith,Bauer,Sekine,Wu} serve as pivotal examples of noncentrosymmetric chiral superconductors.
	Li$_{2}$Pt$_{3}$B exhibits signatures of spin-triplet pairing, including a Pauli-limit-violating upper critical field and evidence of a nodal gap structure \cite{Yuan}. 
	In contrast, Mo$_{3}$Al$_{2}$C displays a more complex interplay of electronic orders, characterized by a polar charge density wave transition at $\sim$ 155 K preceding the superconducting transition at  $\sim$ 9 K \cite{Wu}. This coexistence within a chiral structure suggests a strong coupling between electronic instability, lattice symmetry, and potentially unconventional superconducting pairing. 
	More recently, superconductivity in its isostructural compound W$_{3}$Al$_{2}$C with superconducting transition temperature $T_{c}$ = 7.6 K has been discovered \cite{Ying}. Zero-field and transverse-field muon-spin relaxation/rotation measurements reveal the possible unconventional nature of its superconductivity, as evidenced by the large value of $T_{c}$/$\lambda_{\rm {eff}}^{-2}$ comparable to electron-doped cuprate superconductors \cite{Gupta}. Thus, exploration of superconductivity in this material family is promising to find novel chiral superconductors with exotic physical properties.
	
	In this work, we discovered the superconductivity in W$_{3}$Re$_{2}$C with chiral structure, which is isostructural to Mo$_{3}$Al$_{2}$C. The $T_{c}$ of W$_{3}$Re$_{2}$C is about 6.2 K. Further experimental and theoretical results indicate that cubic W$_{3}$Re$_{2}$C is an intermediately coupled type-II BCS superconductor with full isotropic gap, and its superconductivity mainly originates from the low-frequency softened phonon modes. Density functional theory (DFT) calculations further show that multiple Weyl points are located near the Fermi level ($E_\text{F}$). The coexistence of superconductivity and topological band structure in W$_3$Re$_2$C makes it a promising platform for exploring potential topological superconductivity.

\section{Methods}

\subsection{Sample Synthesis.}
W$_{3}$Re$_{2}$C polycrystals were synthesized via arc melting method. High-purity tungsten powder (99.9 \%), rhenium powder (99.9 \%), and carbon powder (99.9 \%) were used as starting materials. Certain amount of W, Re and C powders were weighed, thoroughly ground in an agate mortar, and pressed into a pellet. This pellet was then arc-melted on a water-cooled copper hearth under an ultra-high-purity argon atmosphere using a tungsten electrode. A zirconium button was employed as an oxygen getter during melting. To ensure homogeneity, the sample was flipped and remelted for 5 -- 6 times. Mass loss after melting was typically less than 1.0 \%.
	
\subsection{Material Characterizations.}
The powder X-ray diffraction (PXRD) was performed using a Bruker D8 X-ray diffractometer with Cu radiation ($\lambda$ = 0.15418 nm) at room temperature. Rietveld refinement of the PXRD pattern was performed using the code TOPAS4.
The composition of W$_{3}$Re$_{2}$C polycrystal was determined by examination of multiple points on the crystals using energy dispersive X-ray spectroscopy (EDX) in in a FEI Nano 450 scanning electron microscope. Electrical transport and heat capacity measurements were carried out in Quantum Design Physical Property Measurement System (PPMS-14T) and magnetization measurements  were carried out in Quantum Design Magnetic Property Measurement System (MPMS3).
	
\subsection{Theoretical Calculations.}
The electronic structures, phonon spectra, and superconducting properties of W$_{3}$Re$_{2}$C were studied based on DFT\cite{DFT-1,DFT-2} and density functional perturbation theory (DFPT) \cite{DFPT-1,DFPT-2} by using the Quantum ESPRESSO package \cite{QE}. The electron-nuclei interactions were described with the RRKJ-type ultrasoft pseudopotentials \cite{USPP} in the Perdew-Burke-Ernzerhof (PBE) \cite{PBE} formalism taken from PSlibrary \cite{PS-1,PS-2}. The kinetic energy cutoff for plane-wave basis and charge densities were set to 60 Ry and 600 Ry, respectively. The Gaussian smearing method with a width of 0.02 Ry was used to broaden the Fermi surface. The maximally localized Wannier functions method \cite{MLW-1,MLW-2} was used to calculate the Fermi surface, which was visualized with the FermiSurfer package \cite{FS}. Considering the large number of atoms in the primitive cell, we chose the 3×3×3 \textbf{q}-point and 18×18×18 \textbf{k}-point meshes for the dynamical matrix and electron-phonon coupling (EPC) calculations due to the limitation of our computational resources (see Fig. S1  for detailed convergence tests of the \textbf{k}- and \textbf{q}-point meshes). The W 5$d$, Re 5$d$ and C 2$p$ orbitals were adopted to construct the tight-binding Hamiltonian by using Wannier90 package \cite{WANNIER90} and the WannierTools package \cite{WANNIERTOOLS} was used to find the Weyl points and calculate the surface states of W$_{3}$Re$_{2}$C.

According to the EPC theory, the Eliashberg spectral function is defined as \cite{A2F}
\begin{equation}
	\alpha^2 F(\omega) = \frac{1}{2\pi N(E_\text{F})} \sum_{\mathbf{q}\nu} \frac{\delta(\omega - \omega_{\mathbf{q}\nu}) \gamma_{\mathbf{q}\nu}}{\hbar \omega_{\mathbf{q}\nu}},
	\label{eq:1}
\end{equation} 
where $N(E_{\rm F})$ is the electronic density of states (DOS) at the $E_{\rm F}$, $\omega_{\mathbf{q}\nu}$ is the frequency of the $\nu$-th phonon mode at the wave vector \textbf{q}, and $\gamma_{\mathbf{q}\nu}$ is the phonon linewidth,
\begin{equation}
	\gamma_{\mathbf{q}\nu} = 2\pi \omega_{\mathbf{q}\nu} \sum_{\mathbf{k}nn'} |g_{\mathbf{k+q}n', \mathbf{k}n}^{\mathbf{q}\nu}|^2 \delta(\varepsilon_{\mathbf{k}n} - E_\text{F}) \delta(\varepsilon_{\mathbf{k+q}n'} - E_\text{F}),
	\label{eq:2}
\end{equation}
in which $g_{\mathbf{k+q}n', \mathbf{k}n}^{\mathbf{q}\nu}$ is the EPC matrix element. With the above parameters, the total EPC constant $\lambda$ can be obtained \cite{A2F},
\begin{equation}
	\lambda = \sum_{\mathbf{q}\nu} \lambda_{\mathbf{q}\nu} = 2\int \frac{\alpha^2 F(\omega)}{\omega} d\omega.
	\label{eq:3}
\end{equation}
Then the superconducting transition temperature $T_\text{c}$ can be determined by substituting the total EPC constant $\lambda$ into the McMillian-Allen-Dynes formula \cite{TC-1,TC-2}
\begin{equation}
	T_c = \frac{\omega_{\mathrm{log}}}{1.2} \exp\left[\frac{-1.04(1+\lambda)}{\lambda(1-0.62\mu^*) - \mu^*}\right].
	\label{eq:4}
\end{equation}
Here $\mu^*$ is a semi-empirical Coulomb screening constant\cite{mu} set to 0.13  in calculation (see Fig. S2 for $T_\text{c}$ calculated with different $\mu^*$ values), while $\omega_{\mathrm{log}}$ is the logarithmic average frequency
\begin{equation}
	\omega_{\mathrm{log}} = \exp\left[\frac{2}{\lambda} \int \frac{d\omega}{\omega} \alpha^2 F(\omega) \ln (\omega) \right].
	\label{eq:5}
\end{equation}

\section{Results and Discussion}
	
\begin{figure}[tbp]
	\centerline{\includegraphics[scale=0.41]{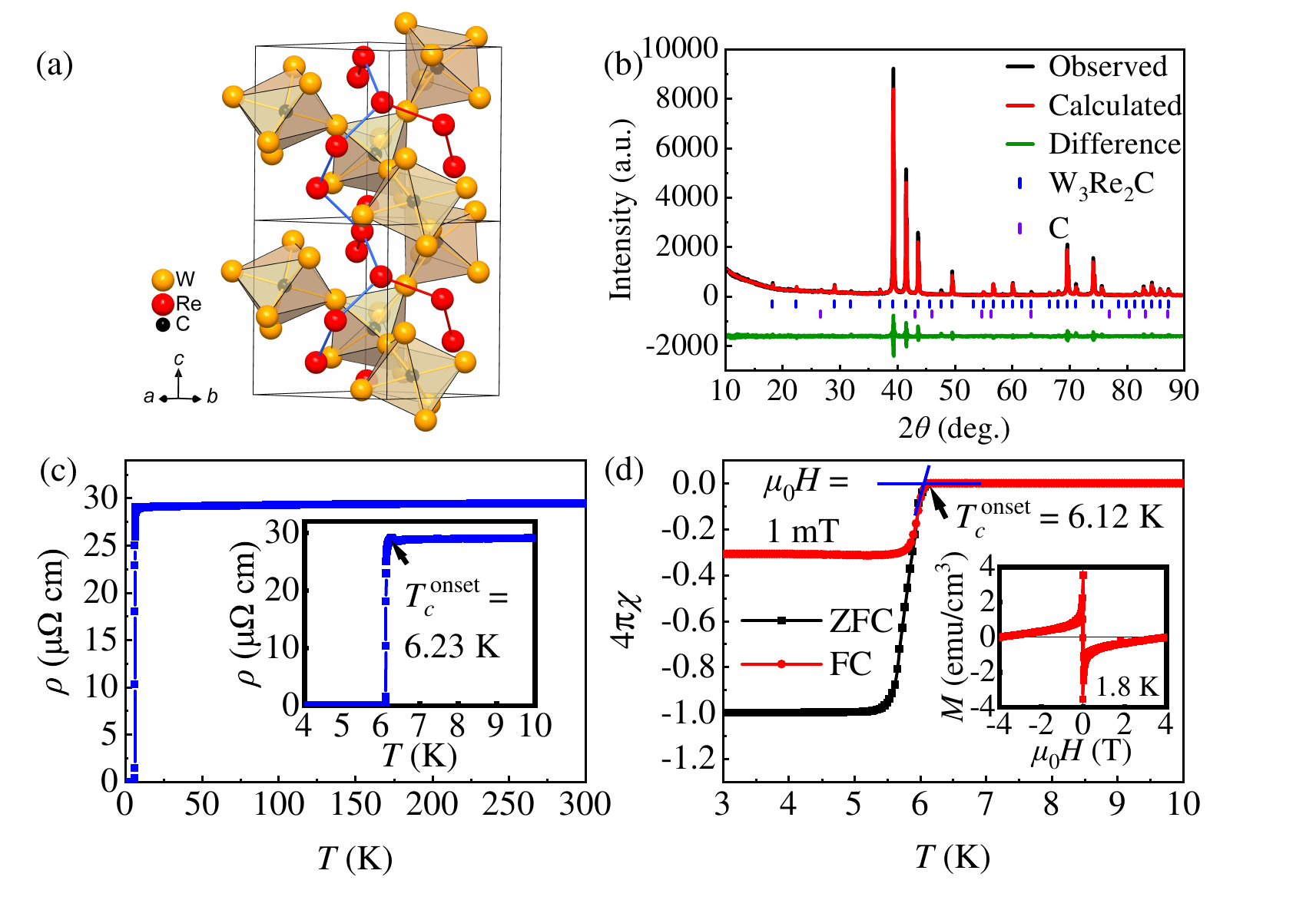}} \vspace*{-0.3cm}
	\caption{
		(a) Crystal structure of W$_{3}$Re$_{2}$C. W, Re, and C is represented by medium orange, big red, and small black balls, respectively. 
		(b) PXRD pattern and Rietveld refinement of W$_{3}$Re$_{2}$C.
		(c) Temperature dependence of $\rho(T)$ at zero field for W$_{3}$Re$_{2}$C  polycrystal. Inset: enlarged view of $\rho(T)$ curve below 10 K. 
		(d) Temperature dependence of 4$\pi\chi(T)$ for W$_{3}$Re$_{2}$C measured in the magnetic field of 1 mT with ZFC and FC modes. Inset: isothermal magnetization loops at $T$ = 1.8 K.}
	\label{fig.1}
\end{figure}

W$_{3}$Re$_{2}$C adopts a cubic structure with the space group $P$4$_{1}$32 (No. 213), corresponding to the \textit{$\beta$}-Mn-type structure \cite{Lawson,Kuzma1963}. 
As shown in Fig. \ref{fig.1}(a), in this structure, each C atom is coordinated with six W atoms to form a distorted W$_{6}$C octahedron. The W$_{6}$C octahedra are connected each other by corner-sharing.
The	Re	atoms reside in the interstitial sites and they exhibit a counterclockwise helical ascent along the $c$ axis. One of such Re-Re bonds is highlighted in blue. The same rotation axis also exists for the W$_{6}$C octahedra.
This crystal structure lacks both inversion and mirror symmetries. Consequently, W$_{3}$Re$_{2}$C has a noncentrosymmetric and chiral structure. The samples with the molar ratios of W : Re : C = 3 : 2 : $x$ were prepared by arc melting method. Based on PXRD patterns and phase analysis,it is found that the purest sample can be obtained with the $x=$ 0.8 and with increasing the C content, and there are significant impurities of W$_2$C appearing in the sample (Fig. S3).
The EDX analysis for $x=$ 0.8 sample yields the atomic ratio of W to Re is 3 : 1.92(1) when setting W as 3, close to chemical formula of W$_{3}$Re$_{2}$C (W : Re = 3 : 2). The corresponding EDX spectrum can be found in Fig. S4.
It is noted that previous study on W$_{3}$Al$_{2}$C found the existence of Al and C deficiencies (W : Al : C = 3 : 1.78 : 0.8) \cite{Ying}, thus there might also be some Re and C deficiencies in W$_{3}$Re$_{2}$C.
Fig. \ref{fig.1}(b) presents the PXRD pattern of this sample, and it can be indexed well by using the structure of W$_{3}$Re$_{2}$C (ICDD, PDF No. 04-003-9331). There is an extra peak located at about 26.6$^\circ$, which can be indexed by elemental carbon with space group $R$-3$m$  (No. 166). This impurity phase may originate from the unreacted starting material of carbon.
Using the Rietveld refinement, the fitted $a$-axial lattice parameter of W$_{3}$Re$_{2}$C is 0.68860(2) nm, in agreement with the results reported in the literature (0.6872 nm) \cite{Lawson}. The detailed refinement results is listed in Table S1.

The temperature-dependent electrical resistivity $\rho(T)$ measured from 300 to 2 K for W$_{3}$Re$_{2}$C polycrystal is shown in Fig. \ref{fig.1}(c) (main panel).
The $\rho(T)$ curve exhibits a poor-metallic behavior with a weak temperature dependence. The residual resistivity ratio RRR (= $\rho$(300 K)/$\rho$(7 K)) is  $\sim$ 1.09.
Such poor metallic behavior of $\rho(T)$ curve with a small RRR value could be due to the grain boundary (GB) scattering in polycrystal sample.
In addition, the existence of defects could also lead to the small RRR value because of electron-impurity scattering. 
It has to be noted that the single crystal of isostructural Mo$_3$Al$_2$C without GB also exhibits a similar behavior \cite{Wu}, implying that it may be an intrinsic behavior of these materials.
Thus, the present $\rho(T)$ data is a higher estimate of intrinsic resistivity because of the existence of GB and/or defect scatterings.
With decreasing temperature further,  a sudden resistivity drop appears at 6.23 K [inset of Fig. \ref{fig.1}(c)], which can be ascribed to a superconducting transition. This superconducting transition shows a very sharp transition width ($\Delta T_{c}<$ 0.15 K).
It is noted that with increasing C content $x$ in the starting materials, both the fitted $a$-axial lattice parameter and the $T_{c}$ exhibit similar non-monotonic behavior but the overall changes are relatively small (Fig. S5). It implies that the C deficiencies may have a minor effect on lattice parameter and superconducting properties.

Fig. \ref{fig.1}(d) presents magnetic susceptibility 4$\pi \chi(T)$ as a function of temperature measured at 1 mT with zero-field-cooled (ZFC) and field-cooled (FC) modes.
The onset superconducting transition temperature $T_{c}^{\rm onset}$ is defined as the intersection between the extrapolated normal-state susceptibility and the steepest-slope line of the diamagnetic signal (blue solid lines). The determined $T_{c}^{\rm onset}$ is about 6.12 K, consistent with the value obtained from $\rho(T)$ curve. 
At $T$ = 1.8 K, after considering demagnetization factor, the superconducting volume fraction estimated from the zero-field-cooling (ZFC) 4$\pi\chi(T)$ curve is close 100 \%, unambiguously confirming the bulk superconductivity of W$_{3}$Re$_{2}$C.
The bifurcation of ZFC and FC 4$\pi \chi(T)$ curves below $T_{c}$ indicates the feature of type-II superconductivity in W$_{3}$Re$_{2}$C and the fluxing pinning effect leads to the smaller value of FC 4$\pi \chi(T)$ when compared to that of ZFC curve. 
Inset of Fig. \ref{fig.1}(d) shows the isothermal magnetization loop $M(\mu_{0}H)$ at 1.8 K, where pronounced hysteresis further confirms the type-II superconductivity of W$_{3}$Re$_{2}$C. 

\begin{figure}[tbp]
	\centerline{\includegraphics[scale=0.43]{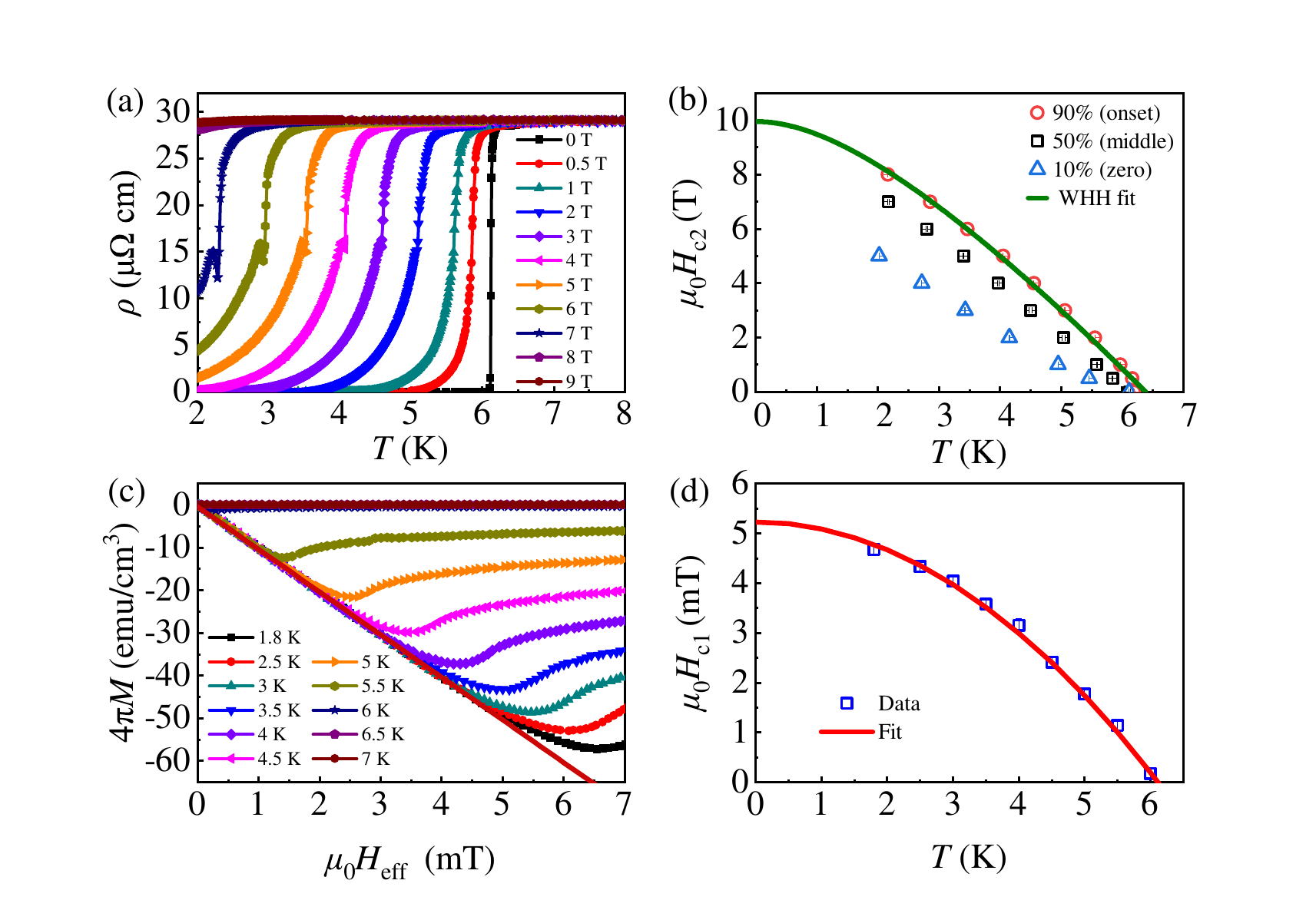}} \vspace*{-0.3cm}
	\caption{
		(a) $\rho(T)$ as a function of temperature at various magnetic fields up to 9 T. 
		(b) Temperature dependence of $\mu_{0}H_{c2}(T)$. The green line represents the fit using the WHH formula. 
		(c) Low-field parts of 4$\pi M(\mu_{0}H)$ curves at various temperatures below $T_{c}$. The red line is the Meissner line. 
		(d) Temperature dependence of $\mu_{0}H_{c1}(T)$. The red line is the fit using the formula $\mu_{0}H_{c1}(T) = \mu_{0}H_{c1}(0)[1 - (T/T_{c})^{2}]$.}
	\label{fig.2}
\end{figure}

Figure \ref{fig.2}(a) shows temperature dependence of $\rho(T)$ under various magnetic fields up to 9 T. With increasing field, the $T_{c}^{\rm onset}$ gradually shifts to lower temperatures. When $\mu_{0}$\textit{H} $\geq$ 8 T, no superconducting transition can be observed above 2 K. 
Interestingly,  the $\rho(T)$ curves under fields show a sharp drop just below $T_c$ then followed by a tail at lower temperature region. In addition, with field larger than 3 T, the dip structure appears in $\rho(T)$ curves.
A similar phenomenon has been observed previously in RbCa$_{2}$Fe$_{4}$As$_{4}$F$_{2}$ and YBa$_{2}$Cu$_{3}$O$_{y}$ single crystals \cite{Xing,Zhu2022,Shibata}. In fact, such anomalous behavior has also been observed in other carbon-containing $\beta$-Mn-type superconductors \cite{Xiao,Kawashima,Zhu2022}.
This phenomenon could be ascribed to either the formation of a vortex slush phase,\cite{Xing} or the presence of precipitates at grain boundaries that create weak links between superconducting grains.\cite{Xiao,Zhu2022}
The upper critical field $\mu_{0}H_{c2}(T)$ was determined using the criteria of 10 \%, 50 \% and 90 \% of the normal-state resistivity $\rho_{n}$ just above the superconducting transition, as shown in Fig. \ref{fig.2}(b). 
It can be seen that three $\mu_{0}H_{c2}(T)$ curves increase with decreasing temperature, and the $\mu_{0}H_{c2,\rm onset}(T)$ curve exhibits a slope $\frac{d\mu_{0}H_{c2}}{dT}\vert_{T_{c}^{\rm onset}(0)}$ = -2.25 T K$^{-1}$, where $T_{c}^{\rm onset}(0)$ is the superconducting onset transition temperature at zero field. 
Furthermore,  the $\mu_{0}H_{c2,\rm onset}(T)$ curve can be fitted well by using the Werthamer-Helfand-Hohenberg (WHH) model (green line) \cite{Werthamer}, yielding a value of  $\mu_{0}H_{c2,\rm onset}(0)$ = 9.95(4) T with $T_{c}$ = 6.38(1) K. This $T_{c}$ is close to that obtained from resistivity measurement. 
Because the Pauli limiting field  $\mu_{0}H_{c2}^{P}(0)$ = 1.84$T_{c}$ = 11.46 T \cite{Aharoni} is larger than $\mu_{0}H_{c2,\rm onset}(0)$, it suggests that the orbital depairing effect should be dominant in W$_{3}$Re$_{2}$C. 
Using the fitted value of $\mu_{0}H_{c2,\rm onset}(0)$, the Ginzburg-Landau (GL) coherence length $\xi_{\rm GL}(0)$ calculated from the equation $\xi_{\rm GL} = \sqrt{\Phi_{0}/2\pi\mu_{0}H_{c2}}$
(where $\Phi_{0} = h/2e$ is the flux quantum) is 5.44(1) nm.

Figure \ref{fig.2}(c) shows the low-field magnetization 4$\pi M(\mu_{0}H_{\rm {eff}})$ curves at various temperatures. 
The effective magnetic field $H_{\rm eff}$ is corrected by considering the demagnetization effect using the formula $H_{\rm eff} =H - 4\pi N_{d}M$, where $N_{d}$ (= 0.23) is the demagnetization factor and $H$ is the external field.
The lower critical field $\mu_{0}H_{c1}$ is defined as the field where the $4\pi M(\mu_{0}H_{\rm eff})$ curves deviate from the linear regime ("Meissner line") with the criterion of $4\pi\Delta  M=$ 1 emu cm$^{-3}$.  
The extracted temperature dependence of $\mu_{0}H_{c1}(T)$ is shown in Fig. \ref{fig.2}(d) and it can be fitted well using the formula $\mu_{0}H_{c1}(T) = \mu_{0}H_{c1}(0)[1-(T/T_{c})^{2}]$ (red line), where  $\mu_{0}H_{c1}(0)$ is zero-temperature lower critical field. The fitted  $\mu_{0}H_{c1}(0)$ is 5.23(2) mT. 
According to the relationship $\mu_{0}H_{c2}(0)/\mu_{0}H_{c1}(0)=2\kappa^{2}/(\rm{ln}\kappa+0.08)$, the calculated GL parameter $\kappa_{\rm GL}$ ($=\lambda_{\rm GL}$/$\xi_{\rm GL}$) is 63.43(3) and the GL penetration depth $\lambda_{\rm GL}$ is 345.1(1) nm, further confirming that W$_{3}$Re$_{2}$C is a type-II superconductor. The zero-temperature thermodynamic critical field $\mu_0H_c(0)$ is estimated to be 111(2) mT using the formula $\mu_0H_c(0) =\mu_0H_{c2}(0)/[\sqrt{2}\kappa(0)]$.

\begin{figure}[tbp]
	\centerline{\includegraphics[scale=0.43]{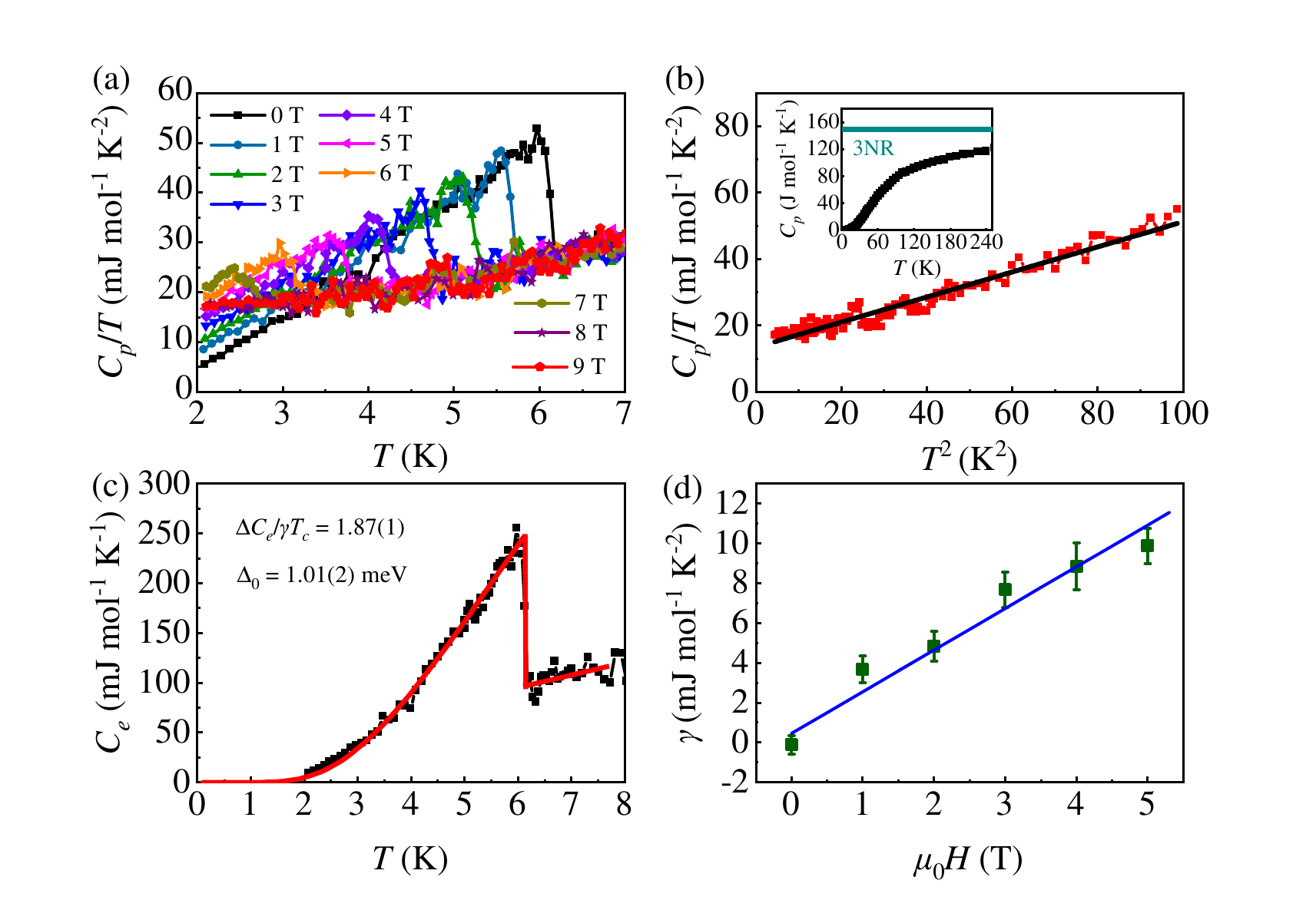}} \vspace*{-0.3cm}
	\caption{
		(a) Temperature dependence of $C_{p}/T$ below 7 K at various fields up to 9 T. 
		(b) Low-temperature specific heat $C_{p}/T$ vs. $T^{2}$ at 9 T. The red solid line represents the fit using the formula $C_{p}(T)/T = \gamma + \beta T^{2}$. Inset shows the zero-field $C_{p}(T)$ curve measured from 2 K to 250 K.
		(c) Electronic specific heat $C_{e}$ as a function of $T$ at zero field. The red solid line is the fit using BCS formula.
		(d) Field dependence of $\gamma$ from 0 T to 5 T. The linear fit is shown as blue solid line.
	}
	\label{fig.3}
\end{figure}

Figure \ref{fig.3}(a) shows the temperature dependence of specific heat  $C_{p}/T$ in the low-temperature region with $\mu_{0}H$ from 0 T to 9 T. It can be seen that there is an obvious jump at $T_{c}$ = 6.17 K for the curve measured at zero field, confirming the bulk superconducting transition in W$_{3}$Re$_{2}$C. The $T_{c}$ is in agreement with the values obtained from the resistivity and magnetization measurements.
With increasing fields, the superconducting jump shifts to lower temperatures and a field above 8 T suppresses the superconducting transition below 2 K. 
As shown inset of Fig. \ref{fig.3}(b), at the high-temperature region, the zero-field $C_{p}(T)$ curve approaches the classical value of 3$NR$ ($\sim$ 149.66 J mol$^{-1}$ K$^{-1}$) predicted by the Dulong-Petit law (green solid line), where $N$ (= 6) is the atomic number and $R$ (= 8.314 J mol$^{-1}$ K$^{-1}$) is the ideal gas constant.
The main panel of Fig. \ref{fig.3}(b) shows the $C_{p}/T$ as a function of $T^{2}$ at 9 T at low-temperature region. It can be fitted using the formula $C_{p}(T)/T = \gamma + \beta T^{2}$, where $\gamma$ is the normal-state electronic specific heat coefficient and $\beta$ is the lattice specific heat coefficient.
The fitted $\gamma$ is 13.49(1) mJ mol$^{-1}$ K$^{-2}$ when the value of $\beta$ is 0.37(1) mJ mol$^{-1}$ K $^{-4}$. 
Correspondingly, the calculated Debye temperature $\Theta_{D}$ is 313(3) K using the equation $\Theta_{\rm D} = (\frac{12 \pi^{4} NR}{5 \beta})^{1/3}$. 
After extracting the phonon contribution, the temperature dependence of electronic specific heat is plotted in Fig. \ref{fig.3}(c). Using the fitted value of $\gamma$, the evaluated specific heat jump $\Delta C_{e}$/$\gamma T$ at $T_{c}$ is $\sim$ 1.87, which is larger than the weak-coupling value 1.43 according to the BCS theory. 
Moreover, the $C_{e}(T)$ curve below $T_{c}$ can be fitted using the BCS formula for the electronic contribution to the specific heat, using the formula $C_{e} = a\exp(-\Delta_{0}/k_{\rm B}T)$, where $k_{\rm B}$ is the Boltzmann constant and $\Delta_{0}$ is the magnitude of the superconducting gap at zero field. 
The fitted value of $\Delta_{0}$ is 1.01(2) meV with the fitting temperature range of 2.08 K -- 5.90 K , which is slightly larger than the weak-coupling BCS theoretical value of 0.93 meV calculated using the formula $2\Delta_{0} = 3.5 k_{\rm B}T_{c}$ with $T_{c}$ = 6.17 K \cite{Tinkham}.
Moreover, the EPC constant $\lambda_{e-ph}$ is obtained from the McMillan equation \cite{McMillan}
\begin{equation}
	\lambda_{e-ph} = \frac{\mu^{*}\ln(\Theta_{D}/1.45T_{c})+1.04}{\ln(\Theta_{D}/1.45T_{c})(1-0.62\mu^{*})-1.04}
	\label{eq:6}
\end{equation}
When assuming the Coulomb pseudopotential $\mu^{*}\sim$ 0.13, the value of $\lambda_{e-ph}$ is determined to be 0.67(2) using $T_{c}$ = 6.12 K and $\Theta_{D}$ = 313(3) K.
All of above results indicate that W$_{3}$Re$_{2}$C is a BCS superconductor with an intermediate coupling strength. It is noted that there are no kinks shown in specific heat curves at high fields in contrast to the behaviors of resistivity curves. Such difference can be understood as following. The resistivity is sensitive to specific defects like the precipitates at grain boundaries, but the specific heat is a bulk-sensitive probe, which may not sensitive to the defects if the volume of defects are very small. Thus, the specific heat measurements may not show the corresponding anomalies. Similar phenomena have been observed in W$_4$IrC$_{1-x}$.\cite{Zhu2022}
Moreover, the field dependent $\gamma(\mu_{0}H)$ can be obtained from the fits using $C_{p}/T=\gamma+\beta T^2$ at superconducting region, which is plotted in Fig. 3(d).
The corresponding fitting curves can be found in Fig. S6.
There is a linear relationship between $\gamma$ and $\mu_{0}H$. This behavior consistently supports that W$_{3}$Re$_{2}$C is a superconductor with full isotropic gap \cite{Nakai}.
It is different from the superconductor with a strongly anisotropic gap or a gap with line node, in which $\gamma$ should follow a nonlinear field dependence, such as $\gamma\propto(\mu_{0}H)^{1/2}$ \cite{Nakai}.

\begin{figure}[t]
	\centering
	\includegraphics[width=0.55\columnwidth]{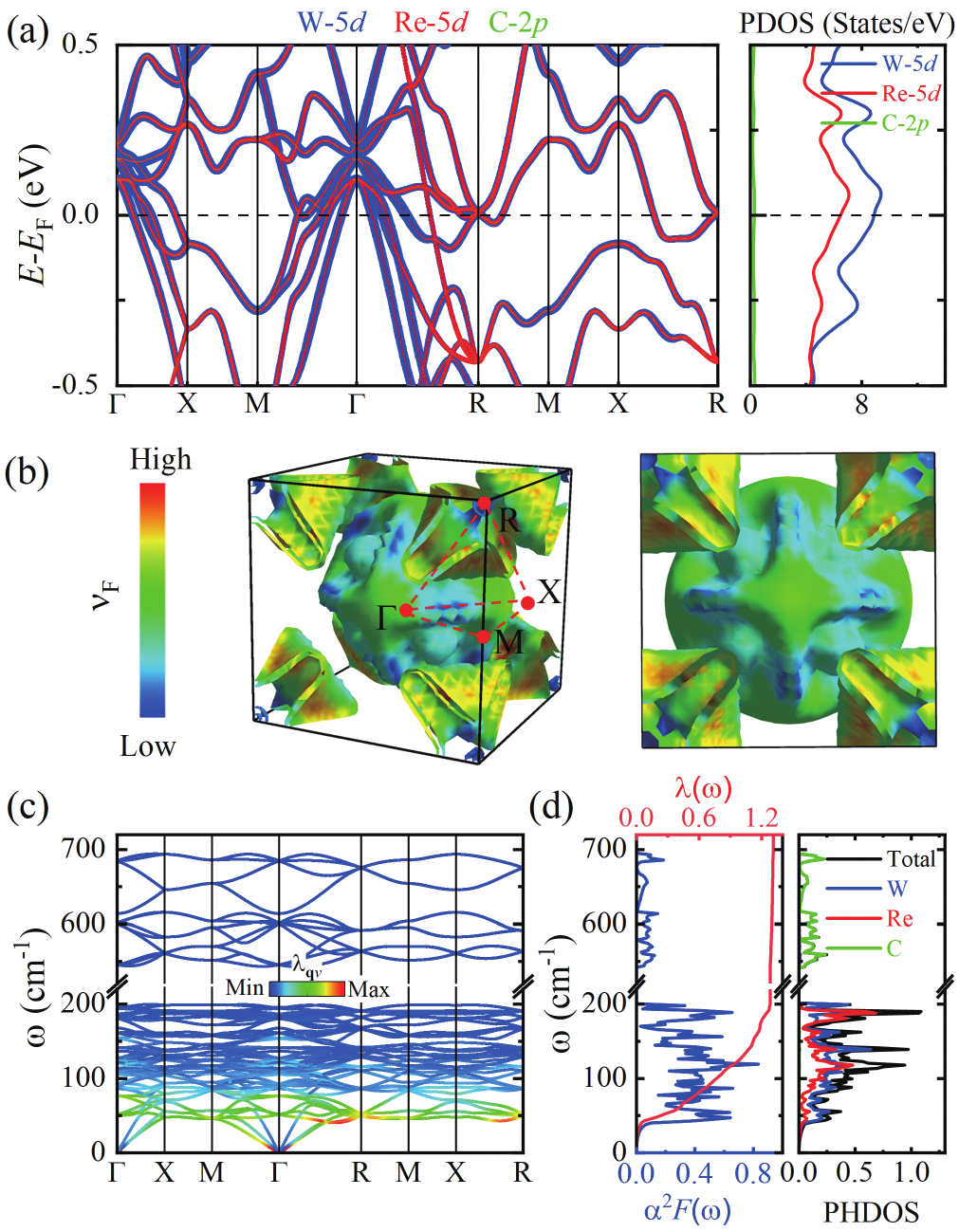}
	\caption{Electronic band structure with orbital weights and PDOS of W$_{3}$Re$_{2}$C. (b) Side and top views of FS sheets with the projection of Fermi velocity. The high symmetry \textbf{k} points are labeled by red dots. (c) Phonon dispersion weighted by the EPC strength $\lambda_{\textbf{q}\nu}$. (d) Eliashberg spectral function $\alpha^2F(\omega)$, frequency-dependent EPC constant $\lambda(\omega)$, and PHDOS.}
	\label{fig.4}
\end{figure}

Figure \ref{fig.4}(a) shows the electronic band structure and the partial density of states (PDOS) of W$_{3}$Re$_{2}$C calculated without the SOC effect. There are multiple bands crossing the $E_{\rm F}$, indicating the good metallic property. Near the R point of the Brillouin zone (BZ), two short flat bands dominated by W 5$d$ and Re 5$d$ electrons show up. According to the PDOS, the electronic states near the $E_{\rm F}$ are mainly contributed by the W 5$d$ and Re 5$d$ orbitals, while the C 2$p$ orbitals are fully occupied and lie far below the $E_{\rm F}$. There is a strong hybridization between the W 5$d$ and Re 5$d$ orbitals, leading to a peak in the PDOS near the $E_{\rm F}$, which may be conducive to the emergence of superconductivity \cite{dos-1}. Figure \ref{fig.4}(b) displays the Fermi surface (FS) sheets weighted by the Fermi velocities, where the red and blue colors represent the highest and zero Fermi velocities, respectively. The FS sheets of W$_{3}$Re$_{2}$C show a pronounced three-dimensional nature, mainly consisting of ball-like hole pockets at the BZ center and the dumpling-shaped electron pockets near the R point.
Considering the large atomic masses of W and Re atoms, we also examined the electronic  structure with the inclusion of SOC and found that the band structure near $E_{\rm F}$ becomes more complex, while the DOS remains nearly unchanged near
 $E_{\rm F}$ but shows 	considerable difference away from  $E_{\rm F}$ (Fig. S7).

\begin{figure}[tbph]
	\centering
	\includegraphics[width=0.565\columnwidth]{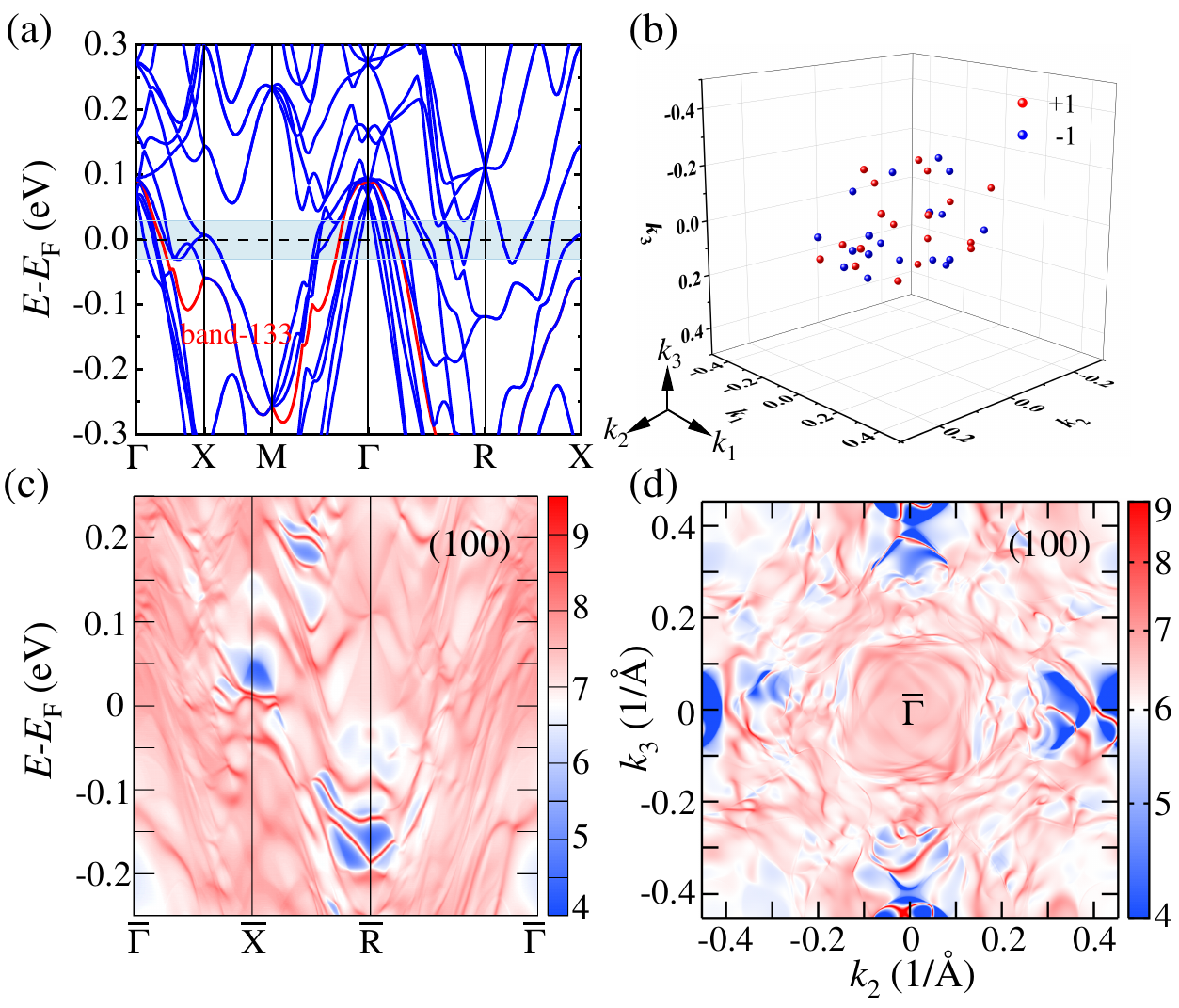}
	\caption{(a) Band structure of W$_{3}$Re$_{2}$C calculated with SOC effect. The red line denotes band 133, which is set as the valence band maximum in the Weyl points calculation. (b) Weyl points plotted in the whole BZ, where red and blue dots possess the topological charge of +1 and –1, respectively. (c) Surface energy bands of W$_{3}$Re$_{2}$C along the high-symmetry paths of projected two-dimensional BZ for the (100) surface terminated by W atoms. (d) Surface spectra of the (100) surface of W$_{3}$Re$_{2}$C at a fixed energy of $E_{\rm F}$ + 20 meV.
		}
	\label{fig.5}
\end{figure}

To better investigate the superconducting properties of W$_{3}$Re$_{2}$C, we performed phonon and EPC calculations. 
The strong EPC in W$_{3}$Re$_{2}$C is further confirmed by the calculated total EPC constant $\lambda$ of 1.27.  Based on the McMillan-Allen-Dynes formula (Equation. ($\ref{eq:4}$)), we obtained the theoretical superconducting $T_\text{c}$ of 10.3 K for W$_{3}$Re$_{2}$C, which is slightly larger than the experimental value (6.12 K). The discrepancy between the theoretical and experimental $T_{\text{c}}$ can be attributed to the following reasons: (1) Neglected SOC effects. Due to computational resource limitation, we are unable to perform  $T_\text{c}$ calculations with the inclusion of SOC, which may have important influence on the EPC and $T_\text{c}$. (2) The existence of grain boundaries. The experimental samples are polycrystalline and contain numerous grain boundaries, which are generally detrimental to superconductivity. (3) Effects of elemental defects. Non-negligible carbon vacancies in the experimental samples may induce disorder and thus decrease  $T_\text{c}$. We also examined the $T_{\text{c}}$ with strong-coupling corrections since $\lambda$ is larger than 1.2, but found only a minor enhancement (Table S2). From the phonon spectrum with the momentum-\textbf{q} and mode-$\nu$ resolved EPC parameter $\lambda_{\textbf{q}\nu}$ (Fig. \ref{fig.4}(c)), we can see that the low-frequency phonon branches below 150 cm$^{-1}$ contribute almost 90$\%$ to the total EPC. There is an obvious softened phonon mode around 50 cm$^{-1}$ along the $\Gamma$-R path, which results in a high peak in the Eliashberg spectral function $\alpha^2F$($\omega$) (Fig. \ref{fig.4}(d)). Combined with the frequency-dependent EPC parameters $\lambda(\omega)$ and the phonon density of states (PHDOS) shown in Fig. \ref{fig.4}(d), the vibrations of W and Re atoms play a dominate role in the superconductivity. In short, we propose that the superconductivity in W$_{3}$Re$_{2}$C belongs to the conventional BCS  type and the EPC is mainly contributed by the couplings of W and Re 5$d$ electronic states with their low-frequency phonons.

In addition to the superconductivity, the breaking of inversion symmetry in W$_{3}$Re$_{2}$C enables the emergence of Weyl points, which may lead to intriguing topological properties and transport phenomena. Since there are many bands near the Fermi level and it is difficult to identify all Weyl points, we chose the band crossings between band 133 and band 134 within a 30 meV energy window near the $E_{\rm F}$ (azury area in Fig. \ref{fig.5}(a)) to confirm the existence of Weyl points. Figure 5(b) shows the positions of these Weyl points in the BZ. In total, 18 pairs of Weyl points are found between band 133 and band 134 near the $E_{\text{F}}$, and their topological charges (chiralities) are listed in Table S3. The chiralities of $\pm 1$ with a total sum of zero confirm their Weyl characteristics. Furthermore, we calculated the surface states of W$_{3}$Re$_{2}$C using a semi-infinite slab based on Green's function method. As shown in Fig. 5(c), there are multiple surface states in the bulk band gaps along the high-symmetry paths of the projected two-dimentional BZ, particularly at the $\overline{\mathrm{X}}$ point around $E_{\text{F}}$ and at the $\overline{\mathrm{R}}$ point at approximately 0.15 eV below $E_{\text{F}}$. The surface spectra taken at at 20 meV above $E_{\text{F}}$ further confirms the existence of these non-trivial topological surface states as shown in Fig. 5(d). Due to the good metallicity and unique chiral structure of W$_{3}$Re$_{2}$C, a pair of Weyl points can exist at different energies and at asymmetric positions in momentum space, which makes it challenging to identify the Fermi arcs connecting a pair of Weyl points with opposite chiralities in Fig. 5(d). The coexistence of superconductivity and Weyl fermions in the chiral W$_{3}$Re$_{2}$C naturally raises the possibility of topological superconductivity, since odd-parity pairing (or mixed-parity pairing) channels are symmetry-allowed and may couple to the topological surface states. More experimental investigations like angle resolved photoemission spectroscopy and scanning tunneling microscope measurements are needed to explore these exotic phenomena in W$_{3}$Re$_{2}$C.

Finally, it is worth comparing to the physical properties of W$_{3}$Re$_{2}$C and the isostructural analogue W$_{3}$Al$_{2}$C. The physical parameters of both materials in superconducting- and normal-states have been listed in Table S4. They exhibit comparable $T_{c}$s and $\kappa_{\rm GL}$s. Although W$_{3}$Re$_{2}$C has a smaller value of $\Delta_{0}$ than that of W$_{3}$Al$_{2}$C, W$_{3}$Re$_{2}$C has a higher DOS at the $E_{\rm F}$, which leads to the larger normal-state $\gamma$ and better metallicity, when compared to W$_{3}$Al$_{2}$C. Notably, because W$_{3}$Re$_{2}$C is predicted to be a Weyl metal (Fig. 5), while the topological property of W$_{3}$Al$_{2}$C has not been reported, it would be interesting to investigate the nontrivial topological property and potential topological superconductivity of W$_{3}$Al$_{2}$C in the future.

\section{Conclusion}

We report a NCS W$_{3}$Re$_{2}$C that crystallizes in the chiral structure, same as Mo$_{3}$Al$_{2}$C. W$_{3}$Re$_{2}$C exhibits a bulk superconductivity with $T_{c}\sim$ 6.2 K.
Further analysis indicates that it is a type-II BCS superconductor with intermediate-coupling strength and full superconducting gap. 
The EPC originates predominantly from interactions between low-energy phonons of W/Re and their 5$d$ electrons. Further studies are worthy to be carried out in order to illustrate the superconducting pairing symmetry and possible nonreciprocal superconducting behaviors in the future.

\section{Supporting Information}

The convergence tests of band structures and phonon spectra with different $\bf{k}$-point and $\bf{q}$-point meshes; the calculated $T_{c}$s with different $\mu^{*}$; PXRD patterns and Rietveld refinements for  W$_{3}$Re$_{2}$C$_{x}$ (${x}$ = 0.8 -- 1); SEM image and EDS/EDX elemental mapping of W$_{3}$Re$_{2}$C polycrystal; the 4$\pi\chi(T)$ curves in the magnetic field of 1 mT with ZFC mode, $a$-axial lattice parameters and $T_{c}$s of W$_{3}$Re$_{2}$C$_{x}$ (${x}$ = 0.8 -- 1); low-temperature $C_{p}/T$ vs. $T^{2}$ under 0 -- 5 T; band structures and DOS calculated with and without SOC effect; a comparison of band structures from DFT and Wannier90; crystal structure parameters and atomic coordinates; the $f_1$ ($f_2$) parameters and the strong-coupling corrected $T_{c}$; the positions and chiralities of the 18 pairs of Weyl points in $\bf{k}$ space; comparison of physical parameters between W$_{3}$Re$_{2}$C and W$_{3}$Al$_{2}$C.

\acknowledgement

This work was supported by the National Key R\&D Program of China (Grants No. 2023YFA1406500, 2022YFA1403800 and 2022YFA1403103), the National Natural Science Foundation of China (Grants No. 12274459, 12174443 and 52572288). Computational resources were provided by the Physical Laboratory of High-Performance Computing at Renmin University of China.

\noindent$^{\dag}$ These authors contributed to this work equally.

\noindent$^{\ddag}$ Present address: Beijing National Laboratory for Condensed Matter Physics and Institute of Physics, Chinese Academy of Sciences, Beijing 100190, China.

\noindent$^{\ast}$ Corresponding authors: K. Liu (kliu@ruc.edu.cn); X. Zhang (zhangxiaobupt@bupt.edu.cn); H. C. Lei (hlei@ruc.edu.cn).

\newpage

\end{document}